# Scaling for Speed on the Water

Joseph L. McCauley
Physics Dept.
University of Houston
Houston, Tx. 77204
jmccauley@uh.edu

## Abstract

Dimensional analysis combined with limited experimental data for the performance of fully submerged propellers have been available since the 1950s. I present two new scale-free hydrodynamic predictions for the relative performance of both submerged and surface-piercing propellers. Larger p/D (pitch to diameter ratio) are more favorable for peak peopeller efficiency and higher speeds; lower p/D are more favorable for carrying loads at low speeds. This conflicts with the common advice to swing as large a diameter as slowly as possible but diameter D has dimensions while p/D is dimensionless, and useful hydrodynamic recommendations must be given in terms of dimensionless variables. For the surface piercing case I compare my scaling predictions with empirical data using a single case in each class as a baseline. One scaling law allows propeller diameter to be predicted from an established baseline where shaft hp and shaft RPM are the variables. The second prediction is inferred from Froude nr. scaling and allows boat speed to be predicted based on shaft horsepower (shp) and weight, given a known baseline in the same class of drag coefficient. I also discuss the existing available data on surface piercing propellers and compare the data with both typical competition data and speed records. In the context of the p/D ratio I discuss the limits on both too high and too low gear ratios. I state for the first time a basic requirement for setting the leading edge camber of surface piercing propellers for optimal acceleration and top speed. I end by using the basic hydrodynamic ideas of circulation conservation and vortex stretching to provide a qualitative picture via the tip vortices of the physics of blade ventilation in surface piercing.

Keywords: Hydrodynamic scaling, tip vortices, propellers, ventilation, Froude nr., supercavitation, circulation conservation, vortex stretching

## 1.Torque, thrust and efficiency of propellers

A hydrofoil or wing experiences lift due to circulation about the foil[1,2]. For small angles of attack relative to the inflow the ratio of lift to drag is large. A propeller has a pitch rather than a fixed attack angle. A propeller blade is a rotating, twisted hydrofoil[3] and also works via circulation with a high thrust to drag ratio. The propeller blade is twisted in order to maintain constant pitch; every point on a blade traces out a helical path as the propeller advances. The thinner the quasi-helicoidal wake, the smaller the drag. Thin blades can make the wake thinner. Speed requires low drag hence thin blades. A helicoid is the surface of constant pitch and describes a propl=eller blade with no camber. Adding camber from leading to trailing edge of the blade takes the blade surface off a helicoid. The observed wake is helicoidal with thickness.

We obtain the standard hydrodynamic predictions for thrust, torque and efficiency from dimensional analysis. Let D be the propeller diameter, n the propeller shaft RPM, U the speed of the boat, and $\rho$ is the water density. The motor's crankshaft RPM is $n_o=n/gr$ where gr is the gear ratio. The thrust developed by the propeller is denoted by $F_T$, and the torque provided by the motor at the propeller shaft is denoted by N. Dimensional analysis[1,2] yields

$$F_T = \rho \kappa_T(J) D^4 n^2 \quad (1)$$

and

$$N = \rho \kappa_N(J) D^5 n^2 \quad (2)$$

where $J=U/nD$ is the dimensionless advance coefficient, $\kappa_T$ is the dimensionless thrust coefficient and $\kappa_N$ is the dimensionless torque coefficient. The physics is hidden in the dimensionless coefficients. These coefficients depend on J and also on other dimensionless parameters: the blade area ratio (b.a.r.)[2]

and blade camber (curvature) are two obvious choices. The b.a.r. is the projection of the blade area onto the plane perpendicular to the propeller shaft and then divided by the area of a disc of diameter D. With $p_{act}=U/n$ the pitch of the (quasi-) helicoidal path traced out by a blade tip it follows that

$$J = \frac{p_{act}}{n}. \quad (3)$$

We can think roughly of $p_{act}$ as the pitch of the wake left by the blade. The relation of blade pitch to wake pitch is $p_{act}=(1-s)p$ where s is called 'slip'. Slip is not the same as efficiency. An auto tire experiences slip when the driver breaks traction. Otherwise the tire has zero slip. A propeller without slip is impossible because of shear in a fluid.

The power supplied by the motor at the propeller shaft (shp, or shaft horsepower) follows from $P_{in}=Nn$ and is

$$P_{in} = \rho \kappa_N(J) D^5 n^3. \quad (4)$$

Typical outboards and stern drives have a shaft hp that is roughly 90% the hp developed by the motor at the crankshaft. The 10% loss is due to the gears and waterpump.

The power delivered to the boat by the propeller is $P_{out}=TU$ or

$$P_{out} = \rho \kappa_T(J) D^4 n^2 U. \qquad (5)$$

The difference between the power input and the power output is due to the wake created by the propeller and gearcase (or propeller and appendages, in the case of inboards). The wake represents pure drag. The mechanical efficiency of a propeller is therefore

$$\eta = \frac{P_{out}}{P_{in}} = \frac{J}{2\pi} \frac{\kappa_T(J)}{\kappa_N(J)}. \qquad (6)$$

Tables of efficiencies, torque and thrust coefficients for fully submerged propellers for different p/L and B.A.R. were produced in the 1950s by Gawn[2,4]. The data were taken for propellers with a an uncambered high pressure blade side and cambered low pressure side, a design that was common for marine propellers in the 1950s and earlier. Note that maximum efficiency corresponds roughly to s≈.05 or about 5% slip. High performance hulls run near maximum efficiency for a given p/L ratio where J=$p_{wake}$/D=(1-s)p/D. We see that the peak efficiency increases as p/D increases. For load-carrying rather than speed J

is lower that for peak efficiency and there lower p/D is advantageous.

In naval architecture or marine engineering the main focus is generally in designing a propeller is often to push a certain load[5], e.g., to propel a ship or other non-planing hull. The marine engineers may fix $F_T$, U, and n and determine D.

In high performance applications the problem is completely different: there is no interest in fixing the thrust to push a certain load, for in $P_{out} = F_T U$ we want large speed U rather than high thrust. The highest thrust is generated by the propeller when the boat is fixed on the trailer or tied to the dock (U=0).

Here's an example of how one can use fig. 1 in a high performance application. This is from a case that I know personally. Suppose that we want to run U=46 mph @ 6600 RPM with a 12:29 gear ratio and a 13" diameter propeller. Then J=1.5. The rig in this case is a classic 1968 Glastron v153, a V-bottom runabout of 550 lb dry weight when empty powered by a 1991 Johnson outboard with 70 shp. The prop used has p/D=19/13=1.46. The curves in fig. 1 don't give us exactly that case. A rough estimate can be obtained if we use a curve with p/L=1.6. We choose curve (a) with the lowest b.a.r. With J= 1.5 we get $\eta$=.72 and $\kappa_T \approx$.08. This yields a thrust T=392 lb and a required power of P=TU/$\eta$=76.2 shp. This prediction is for full submersion of the propeller. The 1991 Johnson 70 is set up for mild surface-piercing (the blade tips exit the water once per revolution). Surface piercing lowers the gearcase drag and reduces the power needed. So by raising the motor and ventilating the propeller we need 6-7 shp less than is predicted for full submersion. We cannot use the curve for p/D=1.4 because in that case $\eta<0$ for J=1.5.

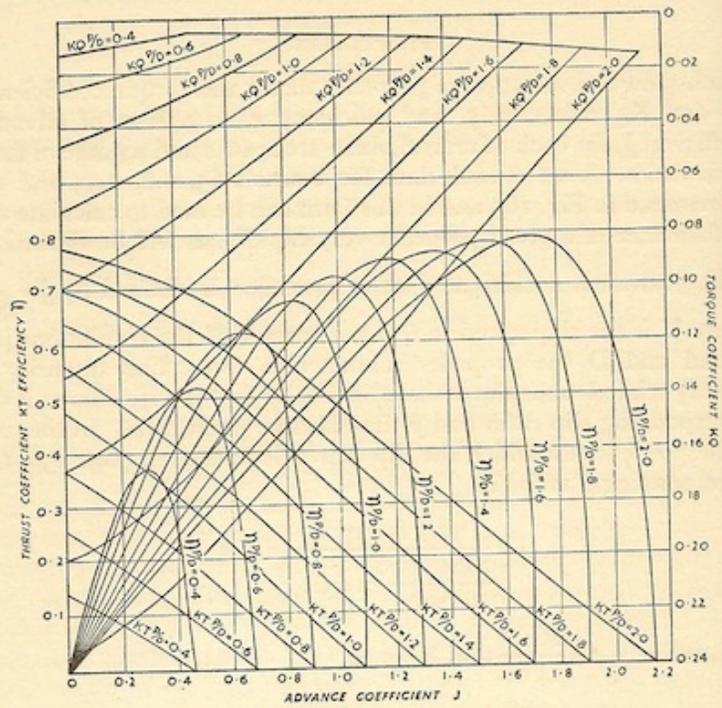

Figs. 165 (a) and (b)

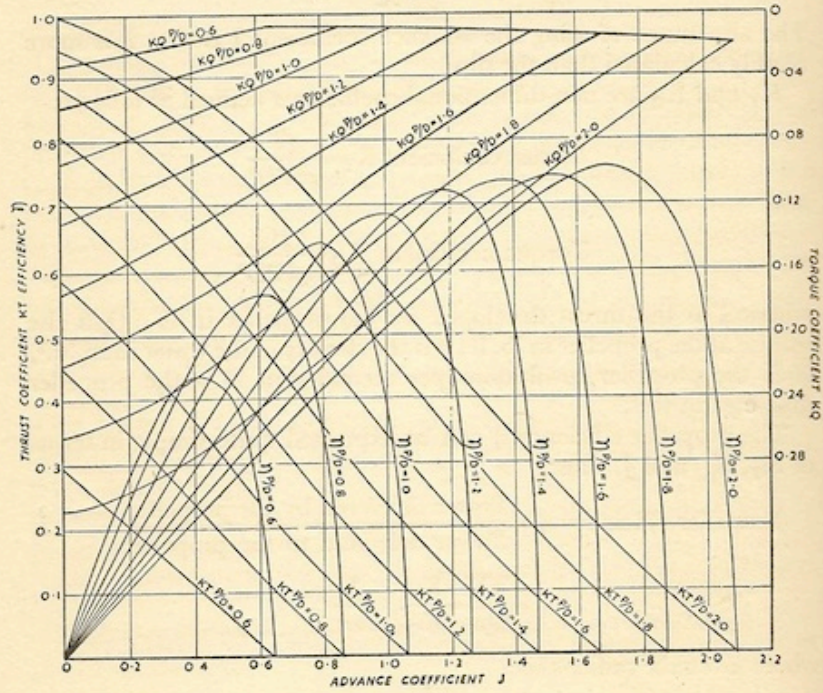

Fig. 1 Efficiency, torque and thrust coefficients for fully submerged propellers.

The graphs tell us nothing about the requirements on the boat, but there is an implicit assumption. The boat weight and bottom (running surface) must be designed so as to permit the speed desired. Minimal requirements are a firm, straight bottom. A boat bottom that flexes permits 'power hooking' and results in speed loss. A bottom with a rocker causes 'porpoising' and speed loss. The transom should meet the bottom as a sharp edge (this reflects the Kutta condition[1] for generation of lift due to a sharp separation of the flow from the bottom). Otherwise there is the tendency of the boat to sink a bit aft and may even generate a rocker effect at the transom.

## 2. Diameter, power and RPM scaling

Consider two different motors with the same gearcase design, hence the same drag characteristics, and roughly the same gear ratios. Then the torque, thrust and drag coefficients should be more or less the same. The same requirement is made on the class of boat to which the power is applied. It follows that

$$\rho \kappa_N(J) = \frac{P_E}{D_E^5 n_E^3} = \frac{P_D}{D_D^5 n_D^3} \qquad (7)$$

so that, given a known baseline, we can use scaling to predict the propeller diameter needed to absorb the power in a different case

$$D_D = D_E \left[ \left( \frac{n_E}{n_D} \right)^3 \frac{P_D}{P_E} \right]^{1/5} \quad (8)$$

where the subscripts D and E refer to two different motors or simply two different gear ratios on the same motor. The torque coefficients depend on J. If the coefficients are different then the above result will be modified by their ratio raised to the power 1/5, and this generally will be near enough to unity to ignore the factor.

We can test this scaling law by using established APBA OPC speed records. I use the 1984 APBA speed record book[5,6] because the records from that era include two sets of two different classes where the same gearcase design was used on motors of different power but with the same gear ratios. I know the propellers and their diameters used in the different racing classes in that era, as well as the motors.

The motor RPM is denoted as $n_o$, shaft RPM is $n=(gr)n_o$. Shaft horsepower (shp), hp measured at the propshaft is 90 % of hp measured at the crankshaft. I know the motor RPM for E and J classes. For the other cases where I do not know the RPM I estimated it from $n_o = 1056 \cdot U / gr \cdot (1-s) \cdot p$, with pitch p given in inches. I know the pitches that were used for kilo records in the various classes. Slip was then taken to be $s \approx .1$ in accordance with high mechanical propeller efficiency. Here's a table including the pitches and diameters used in kilo record runs.

| Class | P(hp) | p(in) | U(mph) | $n_o$(RPM) | gr | $D_{pred}$(in) | $D_{act}$(in) |
|---|---|---|---|---|---|---|---|
| DP | 60 | 27 | 53.89 | 5800 | 13:29 | 13.5 | 13.5 |
| EP | 75 | 25 | 70.56 | 6800 | 15:28 | – | 12 |
| GP | 85 | 30 | 73.19 | 6200 | 13:30 | 14.9 | 14 |
| JP | 150 | 30 | 86.87 | 7000 | 14:28 | 13.9 | 14 |
| VP | 235 | 28 | 84.8 | 6500 | 14:28 | 15.7 | 14.5 |

Table 1

$D_{pred}$ is the diameter predicted by scaling, taking EP Class as the baseline, and $D_{act}$ is the actual diameter used. The worst lack of agreement is for VP Class. Classes GP, JP and VP were splined to run the same Mercury propellers. The VP Class OMC motor was also splined for the Mercury props. Classes DP and EP ran the same OMC props. The props were sold as 13" to 13 1/2" dia. blanks to be thinned, trimmed and recambered.

For the tunnel boat classes I take the SE diameter of 13" as the baseline for predicting the other diameters in table 2.

| Class | P(hp) | p(in) | U(mph) | $n_o$(RPM) | gr | $D_{pred}$(in) | $D_{act}$(in) |
|---|---|---|---|---|---|---|---|
| SD | 60  | 27  | 69.77  | 7400 | 13:29 | 14.7 | 13 |
| SE | 75  | 27  | 91.75  | 7500 | 15:28 | –    | 13 |
| SG | 75  | 27? | 86.87  | 7100 | 15:28 | 13.2 | 13 |
| SJ | 150 | 30  | 102.44 | 8000 | 14:28 | 14.9 | 14 |

Table 2

Had the baseline diameter for SE been taken to be D=12.5", the typical diameter used in competition, then we'd have gotten the predictions of table 2b instead.

| Class | $D_{pred}$ | $D_{act}$ |
|---|---|---|
| SD | 14.1 | 13 |
| SE | –    | 12.5 |
| SG | 12.7 | 13 |
| SJ | 14.3 | 14 |

Table 2b

However, had we taken SJ's L=14" as the baseline we'd have predicted $D_E$=12.2" for SE, which is too small. Typical competition speeds for SE were 80 mph using D=12 ½" and p=23-25".

The records designated as _P (P for 'production' boat) are for stock motors on monohulls (pad V-bottom design) so those records form one class. The records designated S_ (S for 'sport') are for tunnel boats/catamarans. On a catamaran there is nothing in front of the propeller to disturb the inflow. In that case the motors are run in an extreme surface-piercing configuration where not only the entire propeller shaft but also the gearcase hub is above the waterline. For production records the motors were run with the blades less than 50% surface piercing. On the tunnels the amount of submerged blade area is much less.

There are also two current APBA OPC classes that can be compared with each other: SST45 and SST200. Both are tunnel classes with either 1:1 or near 1:1 geared motors with small, very low drag racing gearcases. SST45. Taking SST45 as the standard, table 3 shows the result.

| Class | P(hp) | $n_o$(RPM) | gr | $D_{pred}$(in) | $D_{act}$(in) |
|---|---|---|---|---|---|
| SST45 | 60 | 7000 | 1 | _ | 8.4 |
| SST200 | 250 | 8000 | .93 | 10.73 | 10.25 |

Table 3

In constructing all of the tables I systematically assumed 10 % slip where an assumption was necessary. Next we take on a topic that has been much discussed in the naval architecture literature on the basis of many different wrong predictions.

## 3. Speed, power, weight scaling

We can express the net force on an object due to fluid flowing past it. The free stream speed of the fluid is U. The net drag force on a boat is due to both water and wind. Wind drag is negligible at low speeds. Water drag breaks down into skin friction and form/shape drag. Skin friction is important for boats that plow through the water but matters little for boats with smooth, straight bottoms that ride high and dry. Form drag includes the wake and waves, the main cost of power for high performance hulls. The lift force that causes planing is due to the pressure of the water pushing upward on the boat.

The formula for the net force, including both lift and drag, is derived in hydrodynamics texts[5],

$$\vec{F} = \iint_S P\hat{n}dS + \rho\nu\iint_S \hat{n}\times\vec{\omega}dS \qquad (9)$$

for flow past a fully submerged object. The first term (P is the pressure) represents both the form drag and the lift. The second term ($\vec{\omega}$ is the vorticity) is the skin friction. For boats and gearcases the wave/spray drag at the air-water interface is dominant.

You can get a feeling for form drag and lift (thrust for a propeller) vs. skin friction by splashing a paddle into the water. If you 'knife' the paddle into the water parallel to the blade then it slices in easily. That motion generates pure skin friction. If you instead splash it in at an angle to the blade then eddies are formed on both sides of the paddle and it takes much more effort to move the paddle through the water. That's form drag combined with lift, skin friction is negligible. Water is shoved out of the way by the flat surface of the paddle and flows in behind the edges forming the eddies, or 'backflow'. This example is qualitatively important for surface-piercing propellers, where each blade must either slice or splash into the water each revolution.

We can estimate the friction[6]. With on the order of one sq. ft. of area surface, the flow over the bottom of a boat running 60 mph (with $\nu=10^{-5} ft^2/s$ for water) has a Reynolds nr. $R=UL/\nu \approx 10^7$ so the boundary layer is turbulent. For turbulent flow past a flat plate the skin friction coefficient is $c_f=.072R^{-1/5}$ so the skin friction on a square foot of running surface is $F_f \approx 21$ lb. The total drag equals the total thrust at constant speed U so that the total drag force is $F_D = \eta P_{in}/U$. With $p_{out}=.9(75hp)$ providing the shp the table gives us $\eta \approx .6$ and $F_D \approx 260$ lb $>>21$ lb, so form drag dominates skin friction by more than a factor of 10.

The Froude nr.[1,2]

$$F = U/\sqrt{gd}$$

with d the depth of submersion, measures the kinetic energy vs. gravitational potential energy and comes into play for planing hulls. A ship plows through the water with both skin friction and form drag but a planing hull generates a positive angle of attack by trying to climb up the bow wave. This is the condition for the onset of lift, and lift onset is the precondition for planing. Before lift onset an planing buoyancy carries all of the boat's weight. Planing boats skim over the water with lift carrying essentially all of the weight at moderate Froude nrs. and higher. All high performance boats plane high and dry, the skin friction is very small compared with the form drag (we include interface waves and spray here). Above a certain speed a well-designed high performance boat plans on the water with relatively little submerged surface area and relatively little submeerged depth of the bottom and sides at the transom. The volume of displaced water is on the order of dA where A is the wetted surface area. The volume dA is very small compared with the total boat volume in our case. The smaller the wet surface area and the smaller the submerged depth, the less the form drag and the higher the speed. Racing boats run on a wet surface on the *order of magnitude* of A≈1 $ft^2$ or a few sq, ft. regardless of boat length. Typical lengths can be 12'-16' for tunnels, 30' for unlimited class inboard hydroplanes,  or 40-60' for offshore boats with multiple engines.

This section is motivated by a practical question. If we know how much power is required to push a given boat a given speed, then how will the speed change if we change the power input, the weight w, or both? Crouch, Wyman and others have attempted to answer this question but their results can't be derived either from hydrodynamics or empirically by plotting data. For one thing, the coefficients in their predictions are scale-dependent so the proposed formulae are not scaling laws at all.

Crouch[7,8] proposed the formula $U = c\sqrt{P_{in}/w}$. You can check the formula by plotting $U/\sqrt{P_{in}/w}$. If the formula is correct then the an empirical plot of the numerator vs. the denominator should produce scatter about a flat line (a constant). You will not get that result if you make the plot, so the formula is wrong. It cannot be derived from analyzing empirical data It's also wrong hydrodynamically-seen because the parameter c is not dimensionless. Crouch's formula and various other stabs in the dark are presented in Lord[7]. Wyman[9] proposed that $U=cL^{1/2}(P_{in}/w_d)^{1/3}$ where L represents a wetted length of boat and $w_d$ is the weight of displaced water at speed U. The guesswork behind this formula is not clear. It cannot be derived from a Froude nr. $\sqrt{F} = U/\sqrt{gL}$ and is not correct: (i) it doesn't reflect hydrodynamic scaling since c has dimensions, and (ii) if you divide the lhs by the rhs and plot the result (which should be constant) then you will not obtain scatter about a horizontal line. There is no basis either in theory or data for the proposed formula.

A boat at rest is supported by buoyancy. The boat climbs the bow wave and then planes with a large trim angle. Above planing speed buoyancy becomes negligible if and when the boat's trim angle falls to a few degrees. In that limit dynamic lift acting on the wet surface area carries the full weight of the boat. With a tunnel at high speeds, both water pressure on the sponsons and air-lift, mainly on the tunnel floor enclosed by the sponsons, share the weight.

The dimensionless nr. of interest for the submerged part of the boat is the Froude nr. $F = U/\sqrt{gd}$ where $g$ is the acceleration due to gravity (9.8 m/s²=32ft/s²) and d is the submerged depth of boat.

The drag force on a boat moving at speed U is

$$F_D = \rho c_D A U^2. \qquad (10)$$

We know that drag depends on weight. Writing

$$\rho A U^2 = \rho c_D A F^2 g d = c_D w_d F^2 \qquad (11)$$

where $w_d$ is the weight of the displaced water, we get drag $F_D$ proportional to the weight of the water displaced by the boat. This can be replaced by the boat weight w only if the boat is moving slowly, but not while planing. A necessary but insufficient condition for planing is the Kutta condition, that there is no backflow up the transom: the water must separate cleanly from the bottom at the transom. I have checked this condition with a 15.3' 950 lb runabout (a Glastron v153 powered by a 70 shp Johnson). The water breaks free from the transom at U≈8 mph and the boat planes at U≈14 mph. The bow breaks over to a reduced trim angle at U≥18mph.

When the boat plans then the weight of the displaced water is smaller than boat weight. We assume that the weight of the displaced water is proportional to the boat weight w even when planing. Combining the unknown dimensionless proportionality factor with the drag coefficient, we can write

$$F_D = wc'_D U^2 / gd \quad (12)$$

with a modified dimensionless drag coefficient. Another way to say it is that we scale the drag vs. the weight, we focus on $F_D / w$. We then have as the power required to overcome the drag

$$P_{out} = wc'_D U^3 / gd. \quad (13)$$

This predicts, for comparable boats with comparable set-ups, comparable gear ratios, and comparable submerged depth that

$$U_2 = U_1 \left( \frac{P_2}{P_1} \frac{w_1}{w_2} \right)^{1/3}, \qquad (14)$$

and answers the question 'how much speed should I get for a different rig comparable to one where I know the power, weight and speed'.

Since we've derived the equation for the power transmitted by the propeller to the boat, we must interpret the P's as shaft hp ($P_{in}$). *The formula applies only to classes of motorboats where the efficiency $\eta$, the thrust and torque coefficients are roughly the same.* E.g., fully submerged props on flat bottom fishing boats form one class, reasonably well-ventilated (moderately surface-piercing) props on V-bottoms form another class, and the case of surface-piercing on tunnels forms a third class. Three point hydroplanes form a different class.

We can test this scaling prediction by using known speed records[10,11]. Table 4 shows the predictions of the scaling law vs. the actual speed records, where the EP record was taken as the baseline.

| Class | w(lb) | P(hp) | $U_{pred}$(mph) | $U_{act}$(mph) |
|---|---|---|---|---|
| DP | 800 | 60 | 66 | 53.89 |
| EP | 840 | 75 | – | 70.56 |
| VP | 1500 | 235 | 84.8 | 83.81 |

Table 4

The VP speed prediction is close. The DP prediction is not. Accurate predictions require that the gear ratios are nearly the same. The EP gear ratio is 15:28 =.536, the VP ratio is 1:2 = .50. However, the DP gear ratio is 14:29 = .414. All these motors were run on pad-V-bottom boats and turned 6500-7000 RPM. To understand the effect of gear ratio we take J to be the same for DP and EP because they use exactly the same gearcase housings and propellers. With roughly the same motor RPM but differing only in gear ratio and holding J=constant we find that

$$J = \frac{U_D}{gr_D L_D} = \frac{U_E}{gr_E L_E}. \quad (15)$$

This yields (with $L_D/L_E$=12.5/12=1.04)

$$U_D \approx U_E \frac{gr_D}{gr_E} 1.04 \quad (16)$$

or $U_D \approx 56.68$ mph, about 3 mph faster than the record but much less than the prediction based on the assumption of the same gear ratio.

Classes GP (85 hp Mercury) and JP (150 hp Mercury) are different than the above three classes because the motors ran low water pickups that allowed them to be run as high on the transom as a tunnel with the entire gearcase bullet above the water line. Taking the JP record as the baseline yields table 5

| Class | w(lb) | P(hp) | $U_{pred}$(mph) | $U_{act}$(mph) |
|-------|-------|-------|-----------------|----------------|
| GP    | 920   | 85    | 75.56           | 73.19          |
| JP    | 1075  | 150   | _               | 86.87          |

Table 5

Now for an apparent anamoly: the JP Class Merc 1500 is geared 1:2 whereas the GP Class Merc 850 is geared much lower at 13:30=.43. If we take J=constant for this pair then we'd predict that the speed record for GP should be roughly $U_G \approx U_J(13 \cdot 2/30) = 75.29 \text{mph}$, not much different than in table 5. Why is the DP vs EP prediction so different? We defer the question until below where we discuss gear ratio.

Next come the tunnel ('sport') classes where all records were set using extreme surface-piercing with much less than 50% submergence of the propeller blades. The gearcase bullet drag was eliminated completely by running the bullet above the waterline. The inflow to the propeller is disturbed only by the gearcase skeg, the rudder. Here, we take SE as the baseline for scaling and the results are shown as Table 6.

| Class | w(lb) | P(hp) | $U_{pred}$(mph) | $U_{act}$(mph) |
|-------|-------|-------|-----------------|----------------|
| SD    | 625   | 60    | 87.18           | 69.77          |
| SE    | 675   | 75    | –               | 91.75          |
| SG    | 750   | 85    | 92.79           | 86.87          |
| SJ    | 950   | 150   | 103.34          | 102.44         |

Table 6

Again, Class SD ran with a gear ratio much less than SE so the correct prediction for the SD speed is via holding J=constant,

$$U_D \approx U_E \frac{gr_D}{gr_E} \approx 70 \text{mph} \qquad (17)$$

which is very close to the actual record speed.

If we make the same calculation for SG vs SJ class we get $U_G = U_J(26/30) = 88.8$mph which is also closer to the actual speed record. For pure gear ratio changes at the same motor RPM and prop diameter, holding J=constant provides the more reliable relative performance prediction.

The SST45 and SST200 kilo records were set in the 1990s and are available on the APBA OPC website.

Here's a prediction made by taking SST45 as the baseline

| Class | P(shp) | w(lb) | $U_{pred}$(mph) | $U_{act}$(mph) |
|-------|--------|-------|-----------------|----------------|
| SST45 | 55 | 700 | 97.4 | 87.54 |
| SST200 | 225 | 1150 | – | 128.43 |

Table 7

The prediction is way off because in eqn. 14 we assumed comparable propeller efficiencies. We know that propeller efficiencies are comparable iff. J and p/L are about the same for the classes compared. If we correct eqn. 14 for the difference in efficiencies

$$U_2 = U_1 \left(\frac{\eta_2}{\eta_1}\right)^{1/3} \left(\frac{P_2 w_1}{P_1 w_2}\right)^{1/3} \qquad (14b)$$

and (from fig. 2 and 3 below) ) take $\eta_2/\eta_1 \approx .52/.67 = .78$ then we obtain Table 7b which is very close to the money.

| Class | P(shp) | w(lb) | $U_{pred}$(mph) | $U_{act}$(mph) |
|-------|--------|-------|-----------------|----------------|
| SST45 | 55 | 700 | 89.38 | 87.54 |
| SST200 | 225 | 1150 | – | 128.43 |

Table 7b

It's easy to show that for fixed shp and fixed speed that $p/D \propto \kappa_T^{1/2}(gr)^{-2/5}$. We know from experiment that prop efficiency increases as p/D increases in the range 1<p/D<2. When p/D is too large then prop walk and blowout limit performance. Maybe the high drag limit is reached with the low gear ratio for the DP/SD Class.

## 4. Efficiency of surface piercing propellers

Surface piercing requires as thin a leading edge as material strength will permit without cracking a blade under stress. The thinner the prop the faster it runs. Slam-free entry also helps to avoid leading edge cracks. The farther aft the center of pressure from the leading edge, the less chance of a leading edge crack. The thickness of the trailing edge is less important for performance. Cleavers are relatively thick at the trailing edge. The old and very effective Mercury chopper, as well as the newer Turbo and Stiletto props are relatively thin at the trailing edge.

Hadler and Heckler[12] were the first to collect data on partially submerged propellers[13]. The data were obtained for several different propellers under the following conditions. In some cases U=12 ft/s was held constant with the RPM varied. In other cases the RPM was held constant while the speed was varied from 0 to 12 ft/s. These conditions are very far from those of high performance boating and we will return to them below.

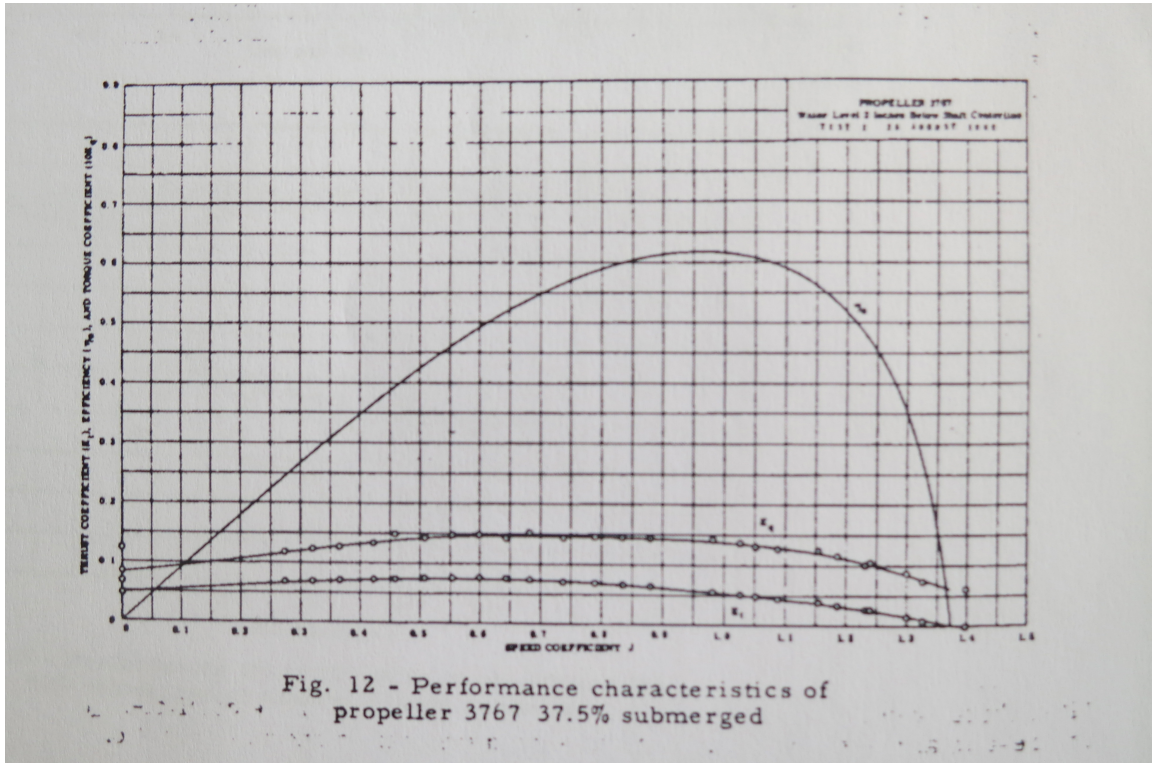

Fig. 2. Efficiency, torque and thrust coefficients for p/L=1.18 with 37.5 % blade submersion.

Figure 3 shows data from Hadler and Heckler for a prop 37.5% submerged, roughly equivalent in motor height to a tunnel boat set up for racing. The pitch to diameter ratio is p/L=1.18, close to an SST45 class tunnel where p/L≈1.17-1.25. The maximum efficiency is still over 60% but the thrust and torque coefficient curves are fairly flat in contrast with the fully-submerged case (in racing, high thrust is not the main factor). The slip at maximum efficiency is about 17%. The data were taken using propellers with blade profiles like cleavers but with an uncambered high pressure side. We have no information about the camber of the low-pressure sides except that all propellers used were of similar camber. Presumably, the props were not cupped. Lack of cup (a very sudden small camber concentrated at the trailing edge, like the flap on an airfoil) reduces the acceleration and low speed performance. The data were taken for props with diameters ranging from 12" to 16".

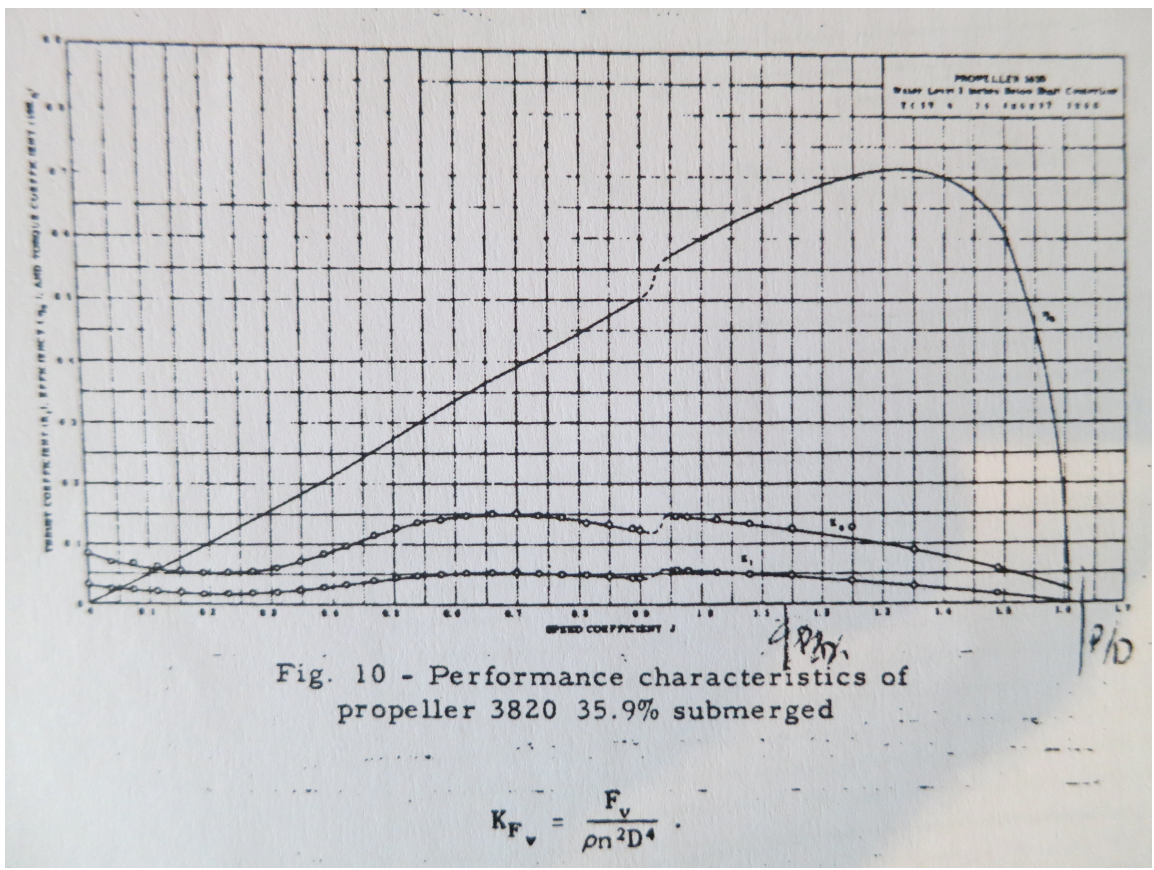

Fig. 10 - Performance characteristics of propeller 3820 35.9% submerged

$$K_{F_v} = \frac{F_v}{\rho n^2 D^4}$$

Fig. 3. Efficiency, torque and thrust coefficients for p/L=1.6 and 35.9% blade submersion.

Figure 5.6 shows data from Hadler and Heckler for a prop 50% submerged, somewhat deeper than the motor height on a tunnel boat. The pitch to diameter ratio is p/D=1.18, close to that of an SST45 class tunnel where p/D≈1.22. The maximum efficiency is 45% @ J=1.03 (compared with .72 @ J=.98 for fully submerged). We can compare with Gawn's graph for p/D=1.2 in ch. 4. The thrust coefficient is a little lower than in the fully submerged case while the torque coefficient is a little higher, hence the 10% lower maximum efficiency for surface piercing. The slip at maximum efficiency is about 14% compared with 18% for fully submerged.

In SST45 Class we have p/D≈1.2 so we'll compare with the data for p/D=1.18 and. With n≈7000 RPM (1:1 gearing) and U≈65 mph (a less-then-winning top speed in short course competition) we have J=1.2. The efficiency predicted by the graph is very low, $\eta$≈.52. I'll comment below why it makes sense not to run at peak efficiency at top speed. Reading off $\kappa_T$≈.05 from the graph we use $F_T = \rho \kappa_T D^4 n^2$ to get a thrust $F_T$=289 lb. Using the formula for shaft hp P=$F_T$U/$\eta$ we get P=97 shp, which is over 75% too large. SST45 motors develop 55 shp @ 6000 RPM and even less at 7000 RPM. Olofsson[15] provided Rolla Propeller data for p/D=1.2: $\eta$≈.64 and $\kappa_T \approx .03$. This gives P≈47 shp, which is much too low. Olofsson's data for p/D=1.24 were taken with a 4 blade prop with large b.a.r. He varied the Froude nr. F and cavitation nr. σ. At his largest Froude nr. F=6 and σ=2.3 he obtained $\eta$≈.71, yielding P≈36 shp, which is also unrealistically low.

We can test the data against another class. Fig. 5.7 shows data from Hadler and Heckler for surface-piercing props with 50% submergence for a pitch to diameter ratio p/D=1.63. We can compare this with Gawn's curve for p/D=1.6. The maximum efficiency is 66% @ J=1.3, in the fully submerged case it was 65% @ J=1.35. The corresponding slips are 20% for surface piercing vs. 15% for fully submerged at maximum efficiency.

We can test this data for SST200 (F1Sport) tunnels powered by 200 shp V-6 outboards running about 110 mph on a short course. The Mercury Optimax 200xs develops at 8000 RPM. With diameter L=10" and pitch p=17" we have p/D=1.7 so we can at best compare with the data for p/D=1.63. Estimating U≈100 mph in short course competition with a 15:17 gear ratio at 8000 RPM yields J≈1.5. With $\kappa_T$≈.02 and η≈.6 we get a thrust $F_T$≈ 155 lb that is way too small and a much too low required power input $P_{in}$≈113 shp. So our comparison with the data produces an overestimate of the required shp for SST45 class and an underestimate for SST200. The Rolla data in Olofsson for p/D=1.6 are h≈.67 and again $\kappa_T$ ≈ .03, yielding P≈151 shp, still far off the mark.

From our standpoint more recent measurements [14,15,16,17] have not improved the data over Hadler and Heckler. To see that we use the Froude nr. which plays a role when the propeller clinbs vertically and pierces the surface of the water. The relevant Froude nr. for propellers is $F=(\text{prop speed})/(g \times \text{depth})^{1/2}$ where $g=32$ ft/s² is the acceleration due to gravity. We can take the depth here to be the depth of submerged diameter $\alpha D$ where $\alpha$ is the fraction of submerged blade. We will take the speed to be the tip speed relative to the water. The Froude nr. of the prop is then $F=W/(\alpha D)^{1/2}$. From part 3.6 the speed of the water over a blade at the tip is $W = nD\sqrt{J^2 + \pi^2}$. With $J \leq 2$ and $\alpha \approx 1/2$ we have $\sqrt{(J^2 + \pi^2)/\alpha} \geq 2.8$ so that making the approximation $F \approx nD/\sqrt{gD}$ (used by Olofsson[16]) underestimates the Froude nr. used by Hadler and Heckler. In the Hadler-Heckler data $U \approx 10$ ft/s, $\alpha \approx 1/2$, and the tip speed is $nD \leq 16$ ft/s so that $F \leq 7$. Even if we underestimate of the Froude by a factor of about 3 by using nr. $F \approx nD/\sqrt{gD}$, then for SST200 and SST45 we get $F \approx 100$ an order of magnitude greater than the Froude nr. in all of the experiments reported (Olofsson used $F \approx 1$-2). We expect that the Froude nr. is important for torque and thrust coefficient measurements, and that measurements made at low F provide results that are inapplicable to high performance boats. Such measurements are generally made in water tunnels with the water merely flowing past a rotating propeller at a controlled speed. Measurements made on boats in motion in open water should be much more reliable. OMC, BRP and Mercury Marine have in the past constructed special gearcases for measuring torque and thrust with the boat in motion on the water but the results are not publically available. Rolla Propeller provides data from water tunnel tests in papers posted on their website, but only for negative trim angles. The problem with low F measurements is that prop walk increases as speed increases (as F increases) and this increases the gearcase drag. Even if the same gearcase would be used in the tests at low F the results will not reflect the increased drag due to the gearcase running cocked at an angle to the direction of motion.

## 5. Circulation conservation for piercing propellers

The scaling laws stated above apply to both fully submerged and surface piercing propellers. The speed records used above were set while running surface piercing ('partially submerged') propellers. A propeller is surface piercing even if only the blade tip exits and reenters the water one time per revolution. Surface piercing changes the physics of blade circulation in interesting ways.

Fully submerged propellers are limited by cavitation: the propeller's tip speed must be low enough that the water doesn't vaporize into bubbles on the blade surface due to the Bernoulli effect. Cavitation leads to a loss of power and to prop damage. Since tip speed goes like $nL/2$, then either the RPM or the diameter must be kept small, or both. Low shaft RPM requires low gearing, but low gearing requires larger prop diameter. The only way to eliminate cavitation and achieve higher speeds is via running the propeller partially submerged. Leaving and re-entering the water once per revolution does not provide enough time for water vaporization and bubble formation at high shaft RPM.

For fully submerged marine propellers the low-pressure blade side is cambered while the high pressure side may not be. For high performance (speeds of 40 to 200 mph, e.g.) both the low and high-pressure sides of the blades should be cambered for surface piercing. The camber of the high-pressure side increases from the leading to the trailing edge just as on the low-pressure side. Blade thickness increases drag by causing a thicker wake and thereby costs speed. For a thin propeller blade the mean camber line[3] is near to the shape of either blade surface. In contrast with airplane wings, where the camber is opposite sign on the two sides of a wing, positive camber on both the high and low pressure blade sides is critical for performance. The airfoil shapes sometimes used in the 1950s are wrong for marine propellers. Airfoils have a large aspect ratio; marine propellers are necessarily of very low aspect ratio. An airfoil section has thickness increasing from the leading edge aft to about mid-chord or less, and then decreases toward a sharp trailing edge. Marine propeller blade camber should never increase and then decrease from leading to trailing edge. Propeller blades can only generate thrust as lift due to circulation. The Kutta condition[3,4] is approximately but well-satisfied on marine propellers even though the trailing edge is not sharp (the trailing edge of a cleaver is generally blunt). Physically, the main point of the Kutta condition is that the fluid leaves the trailing edge sharply, there should be no backflow onto the low pressure side of the blade. The next point is new insofar as publication goes, as are the formulae in sections 2 and 3 above.

In surface piercing the sharp leading edge should be set for 'slam-free' entry. Namely, the ideal camber is roughly equivalent to s=0 over a small chord at the leading edge. The blade angle at the leading edge then produces no thrust locally and very little form drag, and the camber of that small section compensates for the reduced leading edge attack angle. Simple hydrofoil (wing) theory[3] shows how lift is generated on a cambered hydrofoil running even at a small negative attack angle. Correctly set, slam-free camber produces propellers that yield exceptional acceleration with no loss of top speed. Propeller drag is reduced, and camber increases thrust, too much camber will overload the propeller and bog the motor. The application of the slam-free condition depends on the RPM and is applied at the maximum RPM of the motor. For two-stroke racing motors this is typically 7000-9000 RPM. With leading edge camber set correctly, the acceleration from rest can be compared somewhat with the smoothness of an automatic transmission in a car. Except, in the case of the propeller one should never experience a jump like an automatic gear change: the acceleration should be completely smooth from a dead start to top speed with no sudden transition at any speed. When the camber condition is applied to underpowered boats the top speed may be optimized but one loses the smooth transition during acceleration, and may experience supercavitation when trying to plane the boat and even for a period of time afterward.

The main question for physics is: how does a blade recreate circulation as it reenters the water? The circulation conservation theorem[4] demands that the tip vortex must end on the water's surface after both reentry and exit during each revolution. So the wake from a blade that *completely* exits the water is made up of discontinuous sections of a helical tip vortex. Surface piercing propellers are called 'ventilated'. The vortex-helices are the wake, and the ventilation is via the hollow vortex cores ending at the air-water interface.

Marine engineers recognize three different performance regimes for partially submerged propellers[16]. I will deviate from their terminology and will label regimes (ii) and (iii) according to my understanding of vortices. (i) Base ventilated, where the cavity starts on the trailing edge and extends aft to the reentry point of the blade in the water. The propeller develops its highest thrust-to-drag ratio here for a reason explained below. The cavity is simply the hollow-cored (air-filled) tip vortex caused by reentry of the blade, and the vortex stretches as the propeller rotates under water. (ii) Incompletely supercavitating, between base ventilated and fully supercavitating (or 'fully ventilated'). Here, the cavity extends from the air-water interface fore onto the blade's suction side but doesn't extend completely to the leading edge. A sharp transition to the fully ventilated regime would produce a performance drop. (iii) Fully ventilated is the same as supercavitating, an air-filled cavity at atmospheric pressure covers the entire suction side of the blade, starting at the leading edge and ending on the water surface at the points of reentry of the blade. The textbook case[3] of supercavitation is for fully submerged hydrofoils where the cavity extends infinitely far aft of the foil. In the fully ventilated case the thrust is dramatically lower than in the base ventilated case. There must also a sudden performance drop when case (ii) occurs.

A supercavitating/fully ventilated blade operates in a big air cavity and may lose 70-80% of its thrust[16]. This is easy to understand: consider a two dimensional hydrofoil section. The lift coefficient of a supercavitating foil at the same attack angle as a flat plate is only ¼ that of the flat plate[2]. This alone predicts a 75% loss of thrust due to supercavitation. Also, with infinite wingspan the flat plate theoretically has no drag (because there are no tip vortices) whereas a two dimensional supercavitating foil section of infinite wingspan has a finite lift to drag ratio that goes as the inverse of the angle of attack. Supercavitation *dramatically* reduces the lift (thrust coefficient, for a propeller blade) while increasing the drag. In the supercavitating regime the advance ratio J is small with low boat speed and high motor RPM (case (ii) is similar). The first problem is getting the boat to plan, and incomplete supercavitation may continue even after the boat manages with great difficulty to plan. Once on a plan, by suddenly reducing and then increasing the RPM while jerkily changing the motor direction there may be a sudden transition to the base ventilated regime. In the so-called 'base ventilated regime' the blades are not ventilated at all, only the tip vortex is ventilated.

With the propeller properly cambered, and with enough power, supercavitation should not occur in surface piercing applications. Either complete or partial supercavitation destroys performance. Cupping the prop plays a role in avoiding supercavitation, which we may regard as excessive ventilation of the blade. All other things being equal, the blade tip is the most important region because that's where the water speed over the blade is greatest. Cup increases the pressure at the trailing edge by moving the center of pressure aft. For a hydrofoil with a flat mean camber line it's an easy calculation of a standard integral to show that a flap on a hydrofoil can add about 30% more lift. Supercavitation may sometimes be avoided by a dramatic increase in cup, but too much cup will overload the propeller and will correspondingly reduce the top speed. Higher pressure at the trailing edge prevents the formation of the blade-covering cavity. The tendency to supercavitation may also be reduced by correctly cambering the leading edge of the blade. The remedy is to set the leading edge camber so as to avoid both full and partial ventilation, otherwise a boat will not accelerate smoothly from a dead stop and may even fall off a plane in a sharp turn with high motor RPM and low boat speed. The physics of base ventilation can be understood qualitatively by using the ideas of circulation conservation and vortex stretching.

For the case where the blades remain partly submerged from the nine to three o'clock positions the physics of so-called base ventilation is qualitatively simple. The propeller develops adequate thrust and efficiency, peak efficiencies of surface-piercing props are about as high as for fully submerged ones, 60-70%. When a blade section exits the water then the circulation and tip vortex must end at the air-water interface, and the circulation about the submerged part of the blade continues. The submerged section of blade is the vortex core. The circulation ends at the air-water interface on the partly submerged blade, so as a blade rotates from 12 o'clock toward reentry the circulation grows radially over the previously above-water blade section until, at complete blade submersion, the circulation continues off the blade tip with the tip vortex ending on the water's surface at the reentry point. Circulation conservation demands this, and as the propeller blade leaves the interface then the tip vortex is stretched into a helical shape as the propeller rotates and advances. We can regard the entire dynamics of circulation after reentry as vortex stretching. There's a velocity gradient along the blade span, which is the direction of the vorticity in the thin, turbulent boundary layer on the submerged portion of blade. For a fully submerged blade operating efficiently, the only air-filled cavity is simply the blade's wake: there is the hollow tip vortex and also the vortex sheet from the trailing edge (which is unstable, and tends to roll up into a vortex line that combines with the tip vortex, given enough time) just as on a hydrofoil. Air is sucked into the hollow vortex core as the blade reenters the water and the tip vortex stretches, but the air in the vortex core doesn't interfere with the propeller performance *because both sides of the blade are fully wet with full circulation after reentry* in the 'base-ventilated' case. The air-filled tip vortex is itself is a vortex sheet in the form of a tube, the circulation is necessarily the same about the tube as about the blade. The pressure is constant across a tube, and of course the vortex tube will tend to drag the entrained air into solid body rotation. Both blade surfaces are fully wet when submerged and produce normal propeller efficiency. A quantitative approach to this problem would be difficult. The vorticity is confined to the thin, turbulent boundary layer on the blade where we can hope for little more than scaling for quantitative predictions of vortex stretching, if that much.

Here's evidence for base-ventilation and against supercavitation in high performance applications. Fig. 5 shows the low pressure sides of a 12x23 cleaver run on a 75 hp OMC motor on an EP Class race boat at 67 mph with most of the gearcase hub submerged. The exhaust exits the gearcase through the hub, as in fig. 2, but without the removable hub shown on the propeller in fig. 2. That is, the exhaust gas is not confined and is allowed to exit directly onto the entire blade surfaces, water pressure permitting. The restriction of the exhaust deposits to the region near the hub indicate that the blades run wet, not in an air-gas cavity. There is no supercavitation, and this type of exhaust deposit is typical for high performance boats powered by motors with through-hub exhaust (Classes D, E, G, J, and V). Beyond where the carbon deposits end aftward on the hub the gas is mixed with water in the vortex coming off the hub of the propeller. Louis Baumann, observed a case in the 1970s where the low pressure side of the blades emerged from a test run with exhaust streaks. Performance was not up to par. With the SST60 prop in fig. 2 the gearcase hub runs above the waterline. The reason for the slip-on hub on the prop is that the exhaust can ventilate into the vortex core where the blade near the hub protrudes through the water surface and cause supercavitation, known in the pits as 'blow out'.

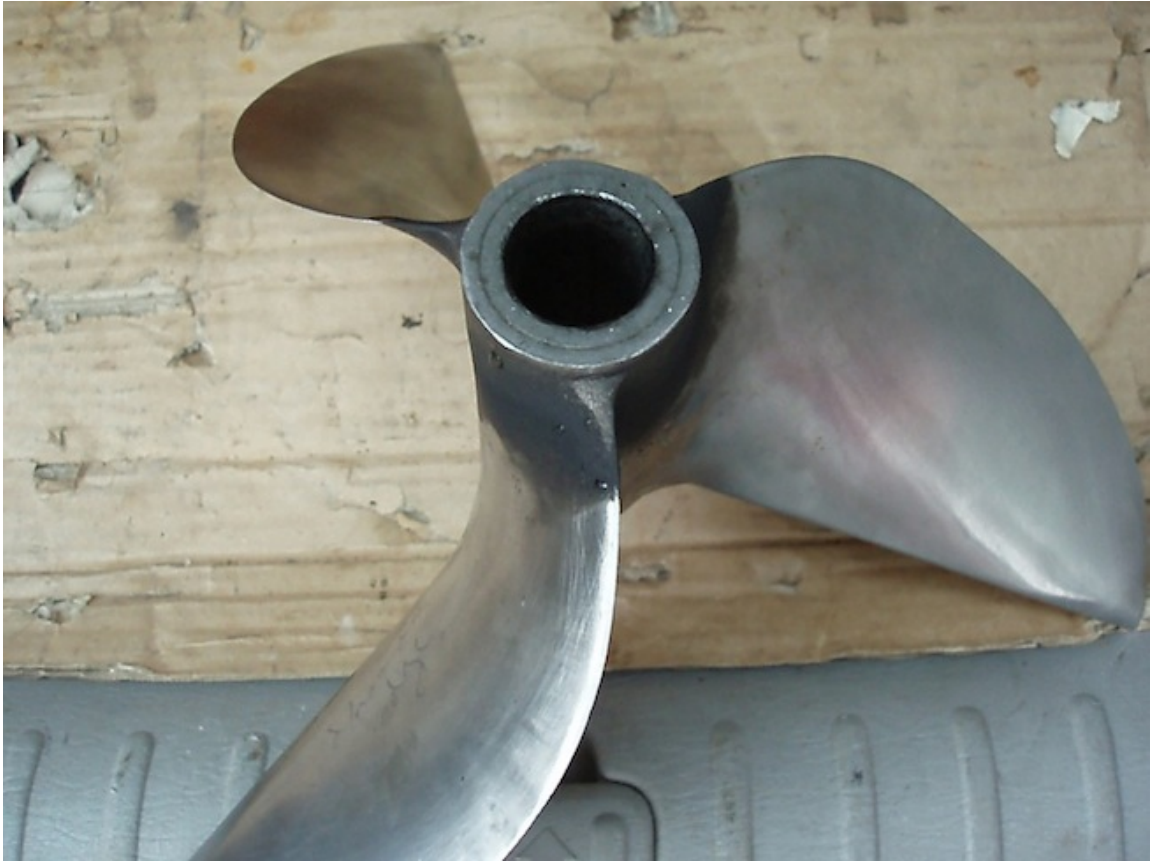

Fig. 4 EP Class prop showing no exhaust deposits where the blades were wet.

The physics of restarting the circulation is somewhat different when the propeller shaft and hub are completely above the water line. Figures 1 and 6 represent this case. Here, the circulation must be started anew with each reentry of a blade. The trailing edge must be submerged to obtain the thrust from circulation. The blade is the vortex core with one end of the vortex ending where the blade pierces the water near the hub. Once started, the tip vortex stretches as the prop rotates and should end at *some* reentry point of the blade on the water's surface. In this case both ends of the vortex provide opportunities for ventilation. We know from circulation conservation[17] that to restart the circulation upon reentry, a trailing vortex of opposite circulation must be shed once per revolution. Upon reentry of the trailing edge the starting vortex should end on the air-water interface near the end of the blade and becomes the tip vortex. This is qualitatively like part of the 'horseshoe vortex' formed by both tips of an airplane wing[3]. Propellers built to run with small submerged fraction of blade area have three or four (or more) blades, so that one blade is always working while the next is reentering. Cupping the trailing edge helps to form the trailing vortex, hence helps to restart the circulation just as dropping the wing flaps helps an airplane to climb.

Fig. 6 shows an F200 Class boat with much less than 50% submersion of the blades. Note the helical wakes coming off the four blades upon exit of each blade. Propeller shaft RPM≈.93x motor RPM =7700, U≈100-110 mph on a short course. The starboard sponson is slightly off the water's surface with the motor trimmed under (the angle of attack of the propeller shaft is negative), reflecting rough water conditions.

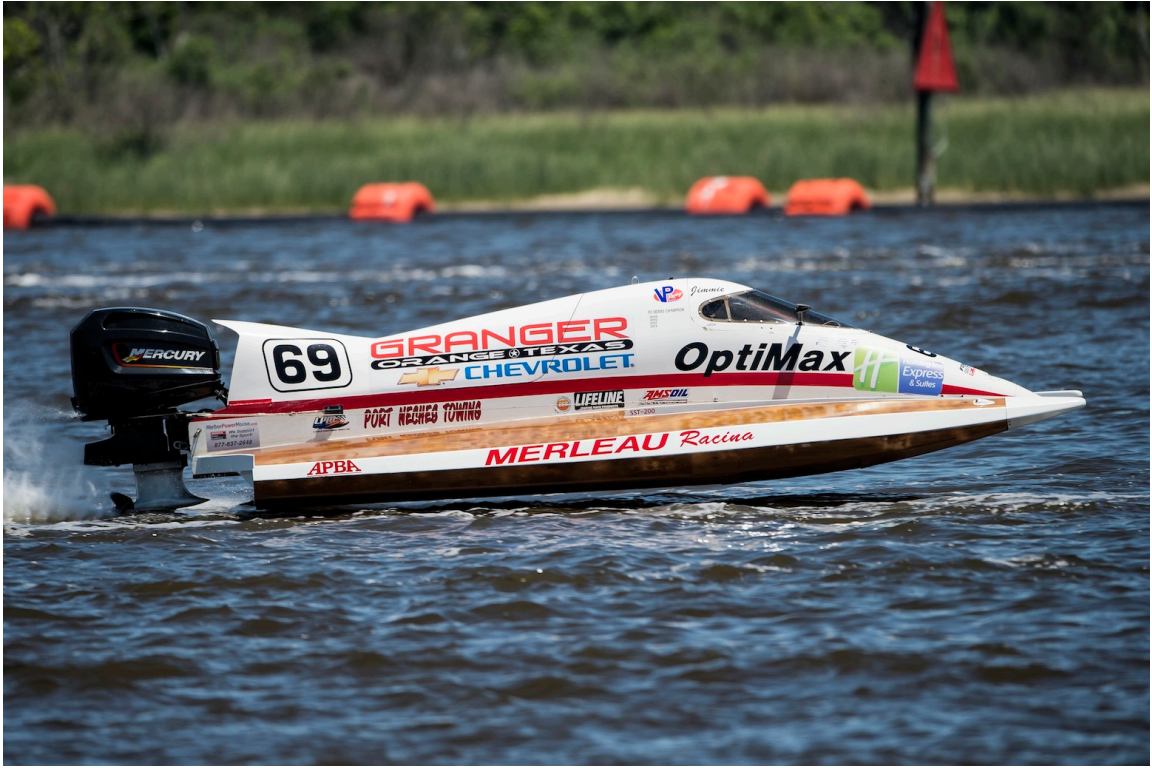

Fig. 5   Surface-piercing propeller. Note the helical wake ( photo by David Alaniz).

## Acknowledgement

I'm very grateful to Edgar Rose and Jim Booe for informative discussions and much encouragement, to Louis Baumann for allowing me to learn propeller work in his shop (Baumann Propeller, Houston, Tx.) in 1978-9, and to Louis Collins, Jeff Titus, and Carlton Callahan for help with some of the data, and to Bodo Bäckmo and John Calley for stimulating discussions. Thanks to Jim Nerstrom for the estimate of hp eaten up by gears and the waterpump. Kevin Bassler encouraged me to write up the scaling results as a paper.